\documentclass[amsmath,floatfix,prb,twocolumn]{revtex4-1}

\usepackage{xcolor}
\usepackage{graphicx}
\usepackage{dcolumn}
\usepackage{bm}
\usepackage{amsfonts}
\usepackage{amsmath}
\usepackage{enumerate}

\newcommand{\Fig} [1]  {Fig.~\ref{#1}}

\begin{document}
\newcommand{\bise}{Bi$_2$Se$_3$}
\newcommand{\sio}{SiO$_2$}
\newcommand{\asio}{a-SiO$_2$}

\title{An \textit{ab initio} investigation of Bi$_2$Se$_3$ topological 
insulator \\
 deposited on amorphous SiO$_2$}

\author{I. S. S. de Oliveira}
\email{igor.oliveira@dfi.ufla.br}
\affiliation{Departamento de F\'isica, Universidade Federal de Lavras,\\
   C.P. 3037, 37200-000, Lavras, MG, Brazil}
\author{W. L. Scopel}
\email{wlscopel@if.uff.br}
\affiliation{Departamento de F\'{\i}sica, Universidade Federal do Esp\'irito 
Santo, Vit\'oria, ES, 29075-910, Brazil}
\affiliation{Departamento de Ci\^encias Exatas, Universidade Federal 
Fluminense, 
Volta Redonda, RJ, 27255-250, Brazil}
\author{R. H. Miwa}
\email{hiroki@infis.ufu.br}
\affiliation{Instituto de F\'isica, Universidade Federal de Uberl\^andia, \\
        C.P. 593, 38400-902, Uberl\^andia, MG, Brazil}
          
\date{\today}
    
\begin{abstract}

We use first-principles simulations to investigate the topological properties of 
Bi$_2$Se$_3$ thin films deposited on amorphous SiO$_2$, \bise/\asio, which is a 
promising substrate for topological insulator (TI) based device applications. 
The \bise\ films are bonded to  \asio\ mediated by van der Waals interactions. 
Upon interaction with the substrate, the  Bi$_2$Se$_3$ topological surface and  
interface states remain present, however the degeneracy between the Dirac-like 
cones is broken. The energy separation  between the two Dirac-like cones 
increases with the number of Bi$_2$Se$_3$ quintuple layers (QLs) deposited on 
the substrate. Such a degeneracy breaking is caused by (i) charge transfer from 
the TI to the substrate and charge redistribution along the Bi$_2$Se$_3$ QLs, 
and (ii) by deformation of the QL in contact with the a-SiO$_2$ substrate. We 
also investigate the role played by oxygen vacancies ($\rm V_O$) on the 
\asio, which  increases the energy splitting between the two  Dirac-like cones. 
Finally, by mapping the electronic structure of \bise/\asio,  we found that  the
\asio\ surface states, even upon the presence of $\rm V_O$,  play a minor role 
on gating the electronic transport properties of \bise.

\end{abstract}

\maketitle

\section{Introduction}

Three-dimensional (3D) topological insulators (TIs) are a recently discovered 
class of materials and have attracted great attention due to their unique 
properties. They are insulators in the bulk phase but present topologically 
protected metallic states at the surface or interface with a trivial 
insulator \cite{RevModPhys.82.3045}. The topological surface 
states (TSSs)  or interface states (TISs) are spin-polarized Dirac 
fermion states, protected by time-reversal symmetry, and in the absence of 
magnetic impurities these states are insensitive to backscattering processes 
induced by time-reversal invariant impurities or defects  
\cite{RevModPhys.82.3045,wrayNatPhys2011,chenScience2010,xuNatPhys2012, 
abdallaPRB2013,tmsPRB2011}. These properties make TI materials suitable for 
spintronics applications \cite{NatMat.11.409.2012}. In particular, Bi$_2$Se$_3$ 
is one of the most investigated 3D-TI, it has a rhombohedral structure composed 
by quintuple layers (QLs) of Se and Bi convalently bonded, and these QLs are 
stacked along the $c$-axis by van der Waals (vdW) interactions 
\cite{NatPhys.5.438, NatPhys.5.398, NatPhys.460.1101}. 

To take advantage of TIs materials for technological applications is necessary 
that the  topologically protected states remain intact upon 
interaction with other materials. Thus, the substrate choice for Bi$_2$Se$_3$ 
growth should be such that it preserves the TIs properties. 
Bi$_2$Se$_3$ has been grown on various substrates, e.g., graphene, Si(111), 
CaF$_2$, CdS, Al$_2$2O$_3$, SiO$_2$, among others \cite{PSSR.7.50.2013}. 
Amorphous phases of dieletric or insulator materials are widely used for 
electronic applications, and an interesting material often used for this purpose 
is the amorphous SiO$_2$ (a-SiO$_2$). Bi$_2$Se$_3$ thin films have been grown on 
a-SiO$_2$ by different experimental technics \cite{doi:10.1021/nl101260j, 
C3NR03032F}, and it has recently been shown to present topologically protected 
surface states 
\cite{steinberg2010surface,apl.104.241606,chang2015simultaneous}. 

To better understand the interaction between Bi$_2$Se$_3$ thin films and the 
a-SiO$_2$ substrate, we perform a first-principles investigation for increasing 
number of Bi$_2$Se$_3$ quintuple layers deposited on a-SiO$_2$, 
(\bise)$_n$/\asio. We find that the \bise\ layers are bonded to the \asio\ 
surface mediated by vdW interactions; where the lattice structure of the 
\bise\ QLs are preserved. At the interface region we have found a net electronic 
charge transfer from the bottom-most \bise\ QL to the \asio\ surface. There is a 
down-shift (up-shift) of the metallic TSSs (TISs), promoting electronic 
transport mediated by electrons (holes) on the surface (interface) \bise\ QLs. 
We have also considered the presence of oxygen vacancies ($\rm V_O$) on the 
surface, \asio[$\rm V_O$],  which is a quite common intrinsic defect. The 
strength of the \bise$\leftrightarrow$\asio[$\rm V_O$] interaction  is the same 
as that of \bise/\asio, as well as the maintenance of the  \bise\ lattice 
structure. Further electronic structure calculations show that the  \asio\ 
surface states, even upon the presence of $\rm V_O$,  play a minor role on 
gating the electronic transport properties of \bise; where the $\rm V_O$ defect 
level is resonant within the valence band of \bise, lying below the 
topologically protected metallic states. 

\section{Computational details}

The calculations are performed based on the density-functional theory (DFT), as 
implemented in the Vienna \textit{ab initio} simulation package (VASP) 
\cite{Kresse}. We use the generalized gradient approximation (GGA), in the form 
proposed by Perdew, Burke and Ernzerhof \cite{PBE}, to describe the 
exchange-correlation functional.  The Kohn-Sham orbitals are expanded in a plane 
wave basis set with an energy cutoff of 400~eV. The electron-ion interactions 
are taken into account using the Projector Augmented Wave (PAW) method 
\cite{Kresse99}. The  Brillouin Zone is sampled according 
to the Monkhorst-Pack 
method \cite{Monkhorst}, using at least a 
2$\times$2$\times$1 mesh. We have also 
used a functional that accounts for dispersion effects, representing van der 
Waals (vdW) forces, according to the method developed by Tkatchenko-Scheffler 
(TS)~\cite{Tkatchenko}, which is implemented on VASP \cite{Bucko}. The inclusion 
of van der Waals forces in the simulations is necessary to obtain the correct 
vdW gap between consecutive QLs \cite{PhysRevB.72.184101}, the interaction between 
the a-SiO$_2$ substrate and Bi$_2$Se$_3$ is also better described with the inclusion of vdW 
interactions.

The amorphous structure was generated through \textit{ab initio} 
molecular dynamics (MD) simulations based on the DFT approach as implemented in 
the VASP code. In Ref. \cite{scopel2008amorphous}, we present details on the 
generation procedure of amorphous SiO$_2$ bulk structure. In order to 
generate the a-SiO$_2$ surface, we have considered a \asio\ slab, where 
the boundary condition perpendicular to the surface plane (z direction) has 
been removed by introducing a vacuum region of 10 \AA. The atomic positions 
have been relaxed until atomic forces were lower than 25~meV/\AA.

\section{Results and Discussion}

\begin{figure}
  \centering
  \includegraphics[clip]{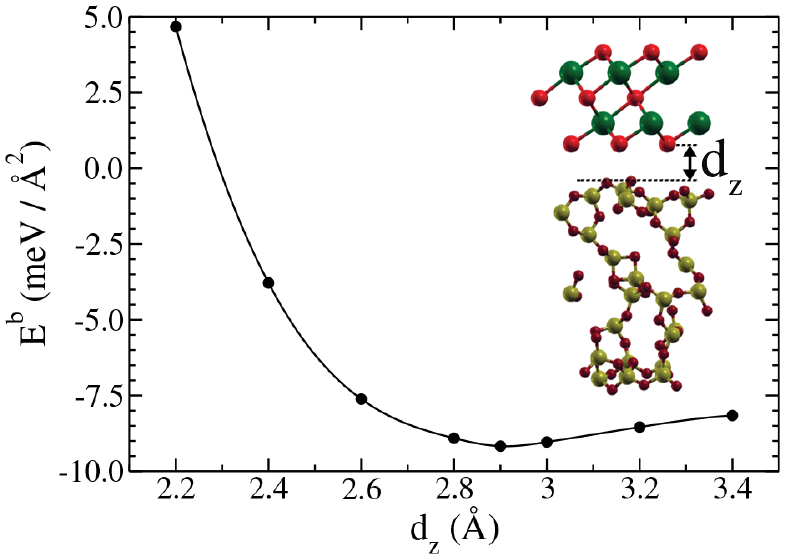}
  \caption{\label{Eads-1QL} Binding energy of 1QL-Bi$_2$Se$_3$/a-SiO$_2$ 
as  a function of separation distance along the perpendicular direction 
($d_z$), 
  as represented in the structure shown in the inset. Green (light red) balls represent
  Bi (Se) atoms, and yellow (dark red) balls represent Si (O) atoms.}
\end{figure}

Our  study starts by calculating the energetic stability and the 
equilibrium geometry of \bise\ on the  \asio\ substrate, \bise/\asio. The  
binding energy ($E^b$) of  \bise/\asio\ is  defined as,
$$
E^b=E[{\rm Bi_2Se_3/a\textendash SiO_2}]-E[{\rm Bi_2Se_3}]-E[{\rm 
a\textendash SiO_2}],
$$
where $E[{\rm Bi_2Se_3}]$ and $E[{\rm a\textendash SiO_2}]$ are the total 
energies of the separated components: a QL of \bise\ and the \asio\ surface; and 
$E[{\rm Bi_2Se_3/a\textendash SiO_2}]$ represents the total energy of the final 
system, a single QL of \bise\ adsorbed on the \asio\ surface, as indicated in 
the inset of \Fig{Eads-1QL}. Our results of $E^b$, as a function of the vertical 
distance $d_z$, are presented in  \Fig{Eads-1QL}. We find $E^{\rm 
b}$=\textendash8.68\,meV/\AA$^2$, for an (averaged) equilibrium distance of 
$d_z$=2.90\,\AA.  As observed for the QLs in the  Bi$_2$Se$_3$ bulk, there is no 
chemical bonding between the  Bi$_2$Se$_3$ and the a-SiO$_2$ substrate, where 
the \bise$\leftrightarrow$\asio\ interaction is ruled by vdW forces.  
Similar picture has been verified for graphene on the \asio\ 
surface, where we found $E^b$=6.3\,eV/\AA$^2$,\cite{miwa2011doping} which is  in 
good agreement with the experimental estimative of Ishigami {\it et al.}, 
6\,meV/\AA$^2$.\cite{ishigamiNanoLett2007}

The absence of chemical bonding between the \bise\ QL and  the \asio\ surface 
has been maintained even upon  the presence of oxygen vacancies ($\rm V_O$) on 
the \asio\ surface (a-\sio[$\rm V_O$]); $\rm V_O$  is a quite common intrinsic 
defect in \sio. We obtained practically the same values of binding energy, 
$E^{\rm b}$=\textendash9.00\,meV/\AA$^2$, and an equilibrium distance $d_z$ of 
 2.96\,\AA. For both systems, \bise/a-\sio\ and /a-\sio[$\rm V_O$], 
the atomic displacements at the interface region are very small, preserving the 
lattice structure of \bise, supporting recent experimental results of  
scanning transmission electron microscopy.\cite{apl.104.241606} Further 
comparisons indicate that, (i) the distance between the substrate and the 
Bi$_2$Se$_3$ QL is approximately 9\% larger than the separation distance between 
consecutive QLs in the Bi$_2$Se$_3$ bulk phase,\cite{PhysRevB.72.184101, 
Oliveira.Nanotechnology2015} and (ii) the  \bise/\asio\ binding energy  is about 
40\% lower (in absolute value) compared with the one between QLs in  \bise\ bulk 
phase. That is, the  \bise$\leftrightarrow$\bise\   interaction is stronger than 
that  between  \bise\ and the \asio\ surface.

\begin{figure}
  \centering
  \includegraphics[clip,width=8.0cm]{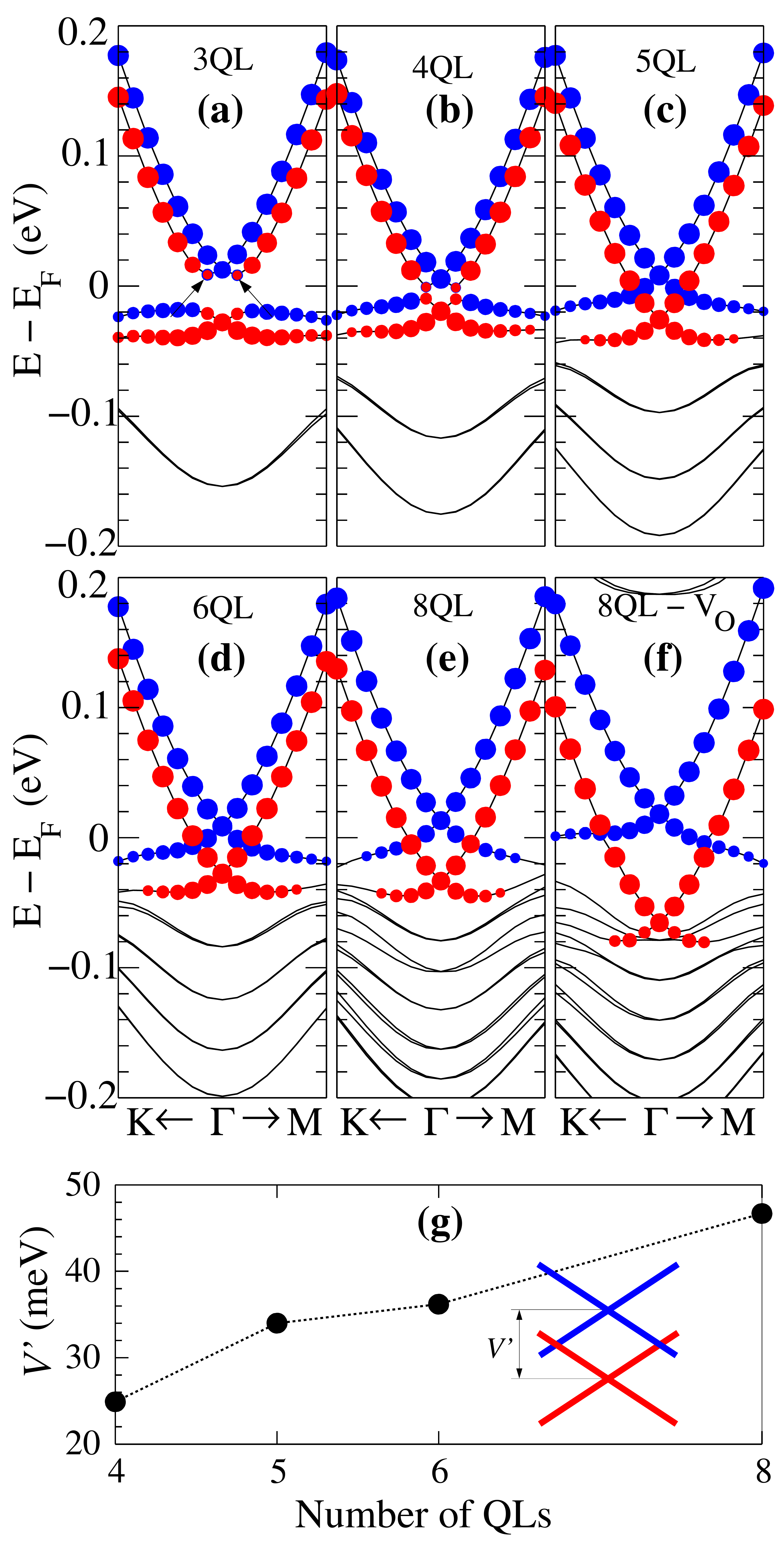}
  \caption{\label{b-proj} \textbf{(a)--(e)} Band structures for $n$QLs-Bi$_2$Se$_3$ 
  deposited on a-SiO$_2$, where $n$ ranges from 3 to 8.
  Blue circles are contributions of Bi $p_z$-orbitals from the bottom-most QL
  (interface with the substrate), and red circles for the top-most QL (vacuum interface). 
  The circles size are proportional to the orbital contribution for each computed k-point. 
  \textbf{(f)} Band structure for (\bise)$_8$/\asio[$\rm V_O$]. 
  \textbf{(g)} Energy gap at the $\Gamma$-point between
  Dirac-like cones from \bise\ topological surface and interface states, as a function 
  of number of QLs deposited on the a-SiO$_2$ substrate.}
\end{figure}

Next we investigate the electronic and topological properties of \bise/\asio, as 
a function of the number of \bise\ QLs ($n$) on the \asio\ surface, 
(\bise)$_n$/\asio. In (\bise)$_n$/\asio, the opposite sides of \bise\ are in 
contact with different environments (with different dielectric constants), {\it 
viz.}: one (top-most) is in contact with a vacuum region, and the other 
(bottom-most) in contact with the \asio\ surface. There is no inversion symmetry 
in (\bise)$_n$/\asio, and thus the energy degeneracy between the edge (QL) 
states can be removed.  Indeed this is what we verify in \Fig{b-proj}, where we 
present the  electronic band structure of  (\bise)$_n$/\asio\ for $n$=3, 4, 5, 
6, and 8. The size of blue (red) circles (in \Fig{b-proj}) is proportional to 
the contribution of the Bi-$p_z$ orbitals of the bottom-(top-)most QL  of the 
(\bise)$_n$. We find that the surface states present an energy gap for $n$=3 and 
4, characterized  by a Rashba-like band splitting near the $\Gamma$ point.  
Those Rashba-like  energy bands  present similar spatial distribution,  and  
spin-texture (not shown) when compared with the ones observed   for  thin films 
of \bise\ on SiC(0001) and InP(111).\cite{zhangNatPhys2010,landolt2014spin} In 
addition, it is worth noting that electronic contributions from the edge  QLs to 
the band apex, indicated by arrows near the $\Gamma$ point in \Fig{b-proj}(a), 
are reduced for $n$=3 and 4. Those states  are delocalized, spreading out along 
the \bise\ layers, as verified for \bise\ on InP(111).\,\cite{landolt2014spin} 
Such a delocalization is supressed upon the formation of gapless Dirac bands for 
$n$ = 5, 6 and 8 [Figs.\ref{b-proj}(c)--\ref{b-proj}(e)], {\it i.e.} the  
topologically protected metallic bands become fully localized on the top-most 
surface, and bottom-most interface QLs of (\bise)$_n$/\asio. 

As shown in Figs.\,\ref{b-proj}(c)--\ref{b-proj}(e), the topological 
surface/interface states (TSSs/TSIs) move downward/upward with respect to the 
Fermi level.  Those results suggest that there is a local $p$-type ($n$-type) 
doping of the edge QLs of \bise. Indeed, based on the  Bader charge analysis 
method,\,\cite{Bader} with the code developed by the Henkelman's 
group,\cite{Henkelman} we find a net charge transfer from the bottom-most \bise\ 
QL to the \asio\ surface atoms, giving rise to  an electrostatic dipole at the 
interface region, as well as an electronic charge density rearrangement along 
the \bise\ slab. In order to provide a measurement of the charge density 
imbalance along the \bise\ layer, we compare the total charge densities at the  
edge QLs of (\bise)$_n$/\asio\ with the ones  of a free standing (\bise)$_n$ 
film. We find that the  charge density reduces by  5.5--6.0$\times 
10^{12}\,e/{\rm cm^2}$ in the bottom-most QL, while it increases by 
0.2--1.0$\times 10^{12}\,e/{\rm cm^2}$ in the top-most QL. Thus, supporting the 
down-shift (up-shift) of the TSSs (TISs) lying on the surface (interface) QL of  
(\bise)$_n$/\asio. 

The separation between the Dirac points [$V^\prime$ in 
\Fig{b-proj}(g)] increases almost linearly with the number of QLs, even one QL 
before the closing of the energy gap, $n$=4. Such a dependence of $V^\prime$, 
with the width ($d$) of (\bise)$_n$, can be attributed to the presence of a net 
electric field ($E^{\rm net}$) along the (\bise)$_n$ film,  $V^\prime\propto 
E^{\rm net} d$.\cite{yazyevPRL2010} Where $E^{\rm net}$ comes from the  charge 
density imbalance discussed above, and the band bending due to the charge 
transfer at the (\bise)$_n$/\sio\ interface. By changing the electronic 
structure at the  \bise/\asio\ interface region, we may have different values of 
$V^\prime$ for a given (\bise)$_n$ width; for instance, the presence of oxygen 
vacancies on the \asio\ surface. As shown in \Fig{b-proj}(f), $V^\prime$ 
increases by 37\,meV, $V^\prime$=47$\rightarrow$84\,meV in 
(\bise)$_8$/\asio[$\rm V_O$].  

Tuning the electron chemical potential within the energy window given by 
$V^\prime$, the energy separation between the opposite edge states of 
\bise/\asio\ will promote  topologically protected electron (hole) currents on 
the top-most surface (bottom-most interface) states of \bise. In addition, the  
up-shift of the interface states in \bise/\asio\ indicates that the scattering 
rate between the TISs and the bulk continuum states will be reduced, while the  
TSSs will face an increase of such a scattering rate.\cite{kim2011surface} 
Further control of the energy positions of those TSSs and TSIs can be done by an 
external electric field. Indeed, top-gate structures has been used to control 
the topological transport properties  in thin films of \bise\ on dielectric 
substrates.\cite{steinberg2010surface,ojeda2012towards, 
liu2015gate,chang2015simultaneous}

\begin{figure}
  \centering
  \includegraphics[clip,width=8.5cm]{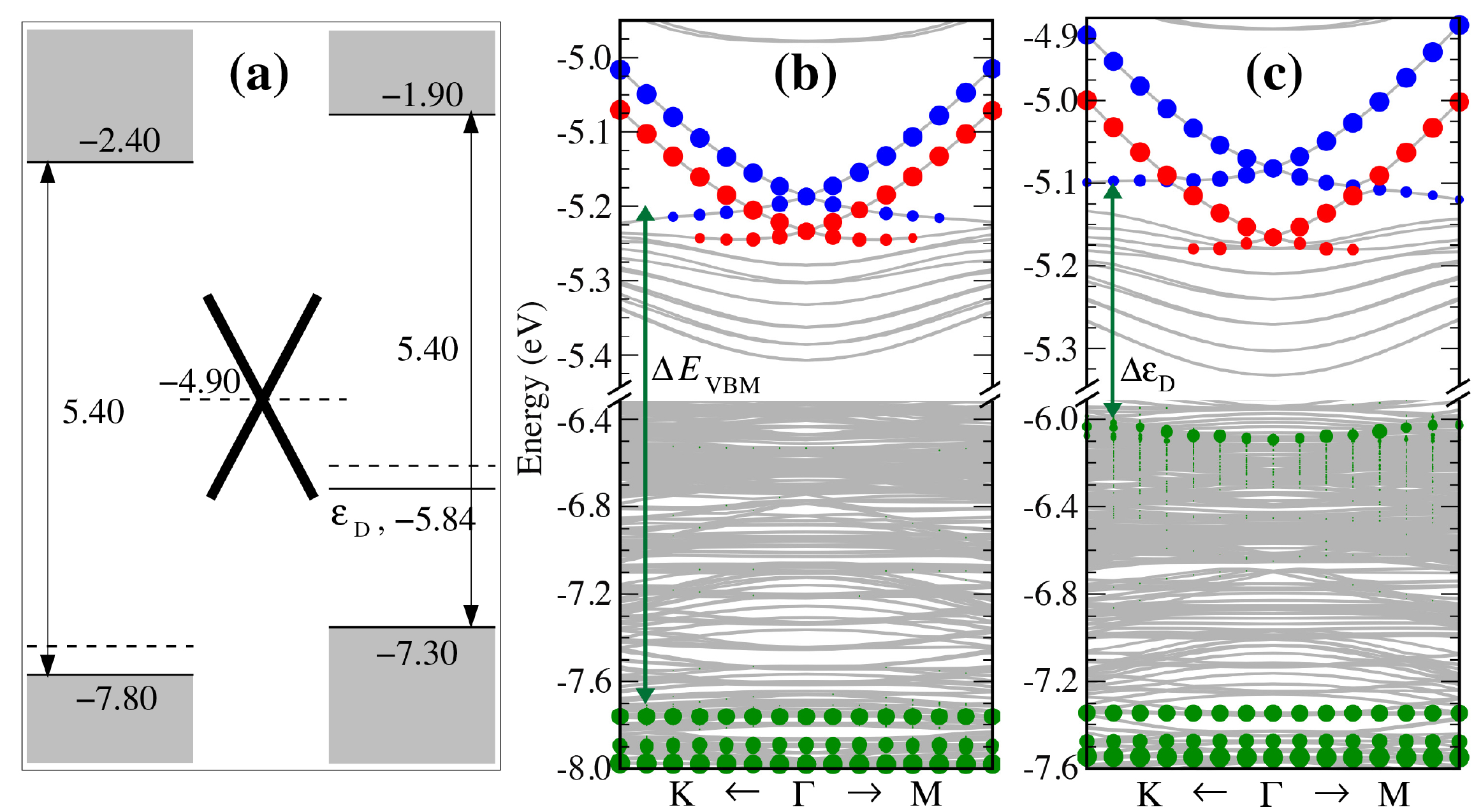}
  \caption{\label{Wfunc} \textbf{(a)} Valence band maximum and conduction band minimum for 
a-SiO$_2$ (left)
  and a-SiO$_2$[$\rm V_O$] (right), dashed black lines represent the Fermi energy, and 
  $\epsilon_D$ 
  the O-vacancy defect level. 
  The Dirac cone for a free standing  (\bise)$_8$ film is represented on the figure's center. 
  \textbf{(b)}  Band structure of (\bise)$_8$/\asio\  and \textbf{(c)}  (\bise)$_8$/\asio[$\rm V_O$].
  Blue (red) circles are Bi p$_z$-orbitals contribution for the bottom-most (top-most) \bise\ QL,
  and green circles the projection of \asio\ orbitals. 
  All energy values are related to the vacuum energy level
   ($E_{\rm vac}=0$).}
\end{figure}

In  Fig.\,\ref{Wfunc}(a), we present the energy positions of the isolated 
systems,  valence band maximum (VBM) and the conduction band minimum (CBM) of 
pristine  \asio\ (left),  and defective  \asio[$\rm V_O$] (right) surfaces. 
Here, we have considered the vacuum level as the energy reference. The doubly 
occupied $\rm V_O$ defect  level ($\varepsilon_{\rm D}$=\textendash5.84\,eV) 
lies at 1.46\,eV above the valence band maximum.\cite{vacancyHSE} In the 
(\bise)$_8$ film [Fig.\,\ref{Wfunc}(a) (center)], the Fermi level is given by 
the crossing on the TSSs; it  lies in the  energy gap of \asio, and above the 
defect level  ($\varepsilon_{\rm D}$) of \asio[$\rm V_O$]. The  position of the 
Fermi energy ($E_{\rm F}$), with respect to the vacuum level, corresponds to the 
work function ($\Phi_0$) of free standing (\bise)$_8$,  $\Phi_0$=4.90\,eV. 
Meanwhile, the work function of the final systems ($\Phi$), (\bise)$_8$/\asio\ 
and /\asio[$\rm V_O$], increases with respect to $\Phi_0$.  Comparing those 
work functions, $\Delta E_{\rm F}=\Phi - \Phi_0$, we can infer the band 
bending   of (\bise)$_8$ upon the formation of the \bise/\asio\ 
interface.\cite{khomyakov2009first} We find positive values of  $\Delta E_{\rm 
F}$,  0.30 and 0.21\,eV for (\bise)$_8$/\asio\ and /\asio[$\rm V_O$], 
respectively, and thus supporting  the  electron transfer from \bise\ to the 
\asio\ surface.

Aiming the development of electronic devices composed by thin films of \bise\  
with  \asio\ as the dielectric gate, it is important to get a picture of the 
energy positions of the TSSs and TISs in \bise/\asio. In Figs.\,\ref{Wfunc}(b) 
and \ref{Wfunc}(c) we present the electronic band structures of the final 
systems, (\bise)$_8$/\asio\ and /\asio[$\rm V_O$], respectively, including the 
orbital projection to the \asio\ surface. Those energy band diagrams show that 
(near the $\Gamma$ point) the TSSs and TISs states lie at about (i) 2.6\,eV 
above the VBM of the \asio\ surface [$\Delta E_{\rm VBM}$ in 
Fig.\,\ref{Wfunc}(b)], and (ii) 0.9\,eV above the $\rm V_O$ defect level 
[$\Delta\varepsilon_{\rm D}$ in Fig.\,\ref{Wfunc}(c)], which is resonant within 
the valence band of \bise. Thus, indicating  that the electronic states of the 
\asio\ surface, even upon the presence of intrinsic defects like $\rm V_O$,    
will play a minor role on gating the electronic transport properties,  
mediated by the topological states,  for \bise\ films.

\section{Conclusions}

We have performed an \textit{ab initio} investigation of Bi$_2$Se$_3$ 
topological insulator deposited on amorphous a-SiO$_2$ substrate. The \bise\ 
layers are  bonded to the \asio\ surface through vdW interactions; preserving 
the lattice  structure of the \bise\ QLs. Topologically protected edge states 
are observed on the surface as well as at the interface layers of \bise, however 
they are no longer degenerated. The TSSs  exhibit an energy-down-shift, followed 
by the up-shift of the TISs. The degeneracy break is caused by the combination 
of two effects: (i) charge transfer from the TI to the substrate and charge 
redistribution along the Bi$_2$Se$_3$ QLs, resulting in electron depletion 
(accumulation) at the closest (furthest) QL from the substrate; and (ii) the 
Bi$_2$Se$_3$ deformation due to the  a-SiO$_2$ 
interaction. Such an energy separation is increased by the presence of $\rm V_O$. 
Finally, by mapping the energy positions of the \bise\ edge states,  the VBM, 
CBM, and the $\rm V_O$ defect level of \asio, we verify  that the \asio\ surface 
states will play a minor role on gating the electronic transport 
properties 
in \bise/\asio\ systems.

\begin{acknowledgments}
 
This work was supported by the Brazilian Nanocarbon Institute of Science and
Technology (INCT/Nanocarbono), and the Brazilian agencies CNPq, FAPES and 
FAPEMIG. The authors also acknowledge the computational support from CENAPAD/SP.

\end{acknowledgments}


%

\end{document}